\documentclass[aps,showpacs,twocolumn,superscriptaddress,nofootinbib,floatfix,section 10pt]{revtex4-2}
\usepackage{amsfonts,amssymb,stmaryrd,latexsym,amsmath,braket}
\usepackage{graphicx,subfigure}
\usepackage{comment}
\usepackage{times}
\usepackage{slashed}
\usepackage{bm}
\usepackage{braket}
\usepackage{appendix}
\usepackage{enumitem}
\usepackage{multirow}
\usepackage{array}
\usepackage{amsmath}
\usepackage[colorlinks=true,backref=true,linktocpage=true,citecolor=blue,urlcolor=blue,linkcolor=blue,pdfpagemode=UseOutlines]{hyperref}
\usepackage[dvipsnames]{xcolor}

\begin{document}
\title{Signal-to-noise problems in fault-tolerant quantum computation
}

\author{Hyunwoo Oh}
\email{hyunwoo.oh5@gmail.com}
\affiliation{Independent Researcher, Tualatin, OR 97062 USA}

\begin{abstract}
Quantum computation offers a promising alternative to classical Monte Carlo methods by naturally avoiding sign problems through unitary evolution.
A common misconception follows from this: because quantum simulation sidesteps the sign problem, it is often assumed to be an essentially perfect substitute for classical sampling. 
This is false---quantum measurements are intrinsically stochastic, and estimating observables often requires a number of samples that grows exponentially with system size, time separation, or inverse temperature, exactly as in the classical signal-to-noise problem. 
In this work, we analyze the origin and structure of signal-to-noise problems in quantum computation, drawing parallels with classical Monte Carlo, and demonstrate that such noise persists even in sign-problem-free regimes. 
We discuss strategies to mitigate it, in particular a quantum generalization of Stein control variates construction, in which unbiased estimators are built from operator commutators $[X,A]$ and their discrete diagonal counterparts, and validate the approach with exactly solvable numerical examples that quantify the achievable variance reduction. 
Our results underscore that variance reduction, not merely sign-problem avoidance, is required to make quantum algorithms practically viable for many-body physics.
\end{abstract}

\date{\today}
\maketitle

\emph{Introduction}---Classical computers have long been relied on Monte Carlo methods as a central tool for studying a wide range of systems in statistical mechanics, quantum many-body physics, and lattice field theory. 
By stochastically sampling over high-dimensional configuration spaces with importance-sampling techniques, Monte Carlo simulations enable the evaluation of thermal and quantum expectation values that would be otherwise intractable to compute directly. 
These methods have produced high-precision results in a variety of domains, including critical phenomena in spin systems, strongly correlated electron systems, thermodynamic properties of quantum gases, and nonperturbative phenomena in lattice gauge theories such as quantum chromodynamics~\cite{Landau_Binder_2021,Gattringer:2010zz,Montvay:1994cy}. 
However, despite these successes, Monte Carlo methods are fundamentally limited by two major obstacles that restrict their applicability in key areas of interest:
\begin{itemize}
    \item \textbf{Sign problems}. In many physically relevant systems, the path-integral weight or sampling probability becomes complex or non-positive. 
    This situation arises frequently in fermionic systems at finite chemical potential, in real-time (Lorentzian) dynamics, and in frustrated quantum spin models~\cite{Troyer:2004ge, deForcrand:2009zkb}.
    Because importance sampling requires a positive-definite weight function to define a probability distribution, such cases must be reformulated using reweighting strategies. 
    This introduces an average phase or sign factor, which typically decays exponentially with system size, inverse temperature, or time extent. 
    As a result, the signal-to-noise ratio of observables deteriorates exponentially, rendering statistically meaningful measurements infeasible beyond small system sizes or short timescales. 
    This is the essence of the sign problem, a notorious barrier to classical simulations of strongly correlated quantum systems.

    \item \textbf{Signal-to-noise problems}. Even in the absence of a sign problem, Monte Carlo estimators can suffer from poor signal-to-noise ratios. 
    This issue arises when the variance of an observable remains large (or even grows) while the expectation value itself becomes small. 
    For example, correlation functions at large Euclidean time separations often decay exponentially with time, while their variances do not decay as rapidly, resulting in exponentially worsening signal-to-noise ratios. 
    In such cases, achieving a fixed relative precision requires an exponentially large number of independent samples. 
    This phenomenon severely limits the ability to extract physical observables such as mass gaps, susceptibilities, or higher cumulants from long-distance behavior.
\end{itemize}

In recent years, quantum computation has emerged as a compelling alternative platform for simulating quantum systems. 
Unlike classical Monte Carlo, quantum computers can implement real-time unitary evolution directly, circumventing the sign problem altogether. 
Algorithms such as the variational quantum eigensolver (VQE)~\cite{Peruzzo:2013bzg}, quantum phase estimation~\cite{Kitaev:1995qy}, and thermal state preparation via purification or quantum imaginary time evolution~\cite{Motta:2019yya} aim to access ground state and finite-temperature physics without recourse to probabilistic reweighting.
These capabilities hold the promise of extending the reach of simulation into regimes long considered inaccessible to classical computation.

However, quantum computers are not exempt from statistical challenges: a widespread misconception holds that, because unitary evolution eliminates the sign problem, a fault-tolerant quantum computer is effectively a perfect simulator free of statistical error. 
This is not the case. Quantum measurement is inherently a stochastic process governed by the Born rule~\cite{Born1926}. 
To estimate the expectation value of an observable, one must perform repeated projective measurements of identically prepared quantum states. 
The central limit theorem still applies, and the precision of the estimator scales as $1/\sqrt{N}$, where $N$ is the number of independent measurements. 
Therefore, quantum computations also face signal-to-noise problems---especially in situations where the observable has large variance or a rapidly decaying signal. 
Examples include out-of-time-order correlators~\cite{Maldacena:2015waa, Swingle:2016var}, long-time dynamics, thermal fluctuations near criticality, and observables dominated by rare events or exponentially suppressed amplitudes. 
In such settings, the number of samples required for accurate estimation can again grow prohibitively large, compromising the practical utility of quantum algorithms.

In this paper, we present a unified analysis of the signal-to-noise problem as it manifests in both classical and quantum computation. 
By exploring common mathematical structures underlying stochastic estimators in both frameworks, we clarify the origins and limitations of measurement-induced noise, and investigate techniques---such as a quantum generalization of Stein control variates---that aim to reduce statistical uncertainty. 
We validate this approach with exactly solvable numerical examples that quantify the achievable variance reduction, and outline how variational, circuit-based optimization could extend the method to larger, non-integrable systems. 
We emphasize that while quantum computation avoids the sign problem, the signal-to-noise problem remains a significant and often underestimated challenge that must be addressed to fully realize the potential of quantum simulation.

\emph{Signal-to-noise problem}---The signal-to-noise ratio (SNR) of an observable $O$ is defined by
\begin{equation}
    {\rm SNR}(O) = \frac{\langle O \rangle}{\sqrt{{\rm Var}(O)}},
\end{equation}
where 
\begin{equation}
    {\rm Var}(O) = \langle O^2 \rangle - \langle O \rangle ^2.
\end{equation}
When ${\rm SNR}(O) \ll 1$, individual measurements of $O$ fluctuate so strongly relative to its mean that an extremely large number of samples $N$ is required to control the standard error,
\begin{equation}
    \sigma_{\bar O} = \sqrt{\frac{{\rm Var}(O)}{N}} = \frac{\langle O \rangle}{{\rm SNR}(O) \sqrt{N}}.
\end{equation}
To achieve a fixed relative uncertainty $\sigma_{\bar O}/\langle O \rangle \approx \epsilon$, one needs
\begin{equation}
    N \approx \frac{1} {\left( \epsilon \cdot {\rm SNR}(O)\right)^2}.
\end{equation}
Thus a vanishing SNR directly translates into an exponential or power-law explosion in computational cost.

By the Born rule, a projective measurement of a state $| \psi \rangle$ onto an orthonormal basis $\{ |\psi_i \rangle \}$ yields outcome $i$ with probability
\begin{equation}
    P(i) = | \langle \psi_i | \psi \rangle|^2.
\end{equation}
Each trial is independent, so estimating $\langle O \rangle$ from repeated measurements is a stochastic process with variance
\begin{equation}
    {\rm Var}(O) = \langle O^2 \rangle - \langle O \rangle ^2,
\end{equation}
and estimation error
\begin{equation}
    \sigma_{\bar{O}} = \sqrt{{\rm Var}(O)/N}
\end{equation}
with $N$ samples.

As an illustrative example, consider a qubit with basis states $\{ |0\rangle , |1\rangle \}$ and an observable
\begin{equation}
    \hat{O} = | 0 \rangle \langle 0 |,
\end{equation}
so its eigenvalues are $\{0, 1\}$. Prepare the pure state
\begin{equation}
    |\psi(p)\rangle = \sqrt{p} | 0 \rangle + \sqrt{1-p} | 1 \rangle,
\end{equation}
with $p \in [0,1] $. Then
\begin{equation}
\begin{aligned}
    \langle O \rangle &= p, \\
    {\rm Var}(O) & = p(1-p).
\end{aligned}
\end{equation}
For $p \ll 1$, one finds
\begin{equation}
    {\rm SNR}(O) \approx \sqrt{p}.
\end{equation}
After $N$ trials, the relative error is
\begin{equation}
    \frac{\sigma_{\bar O}}{\langle O \rangle} = \frac{1} {{\rm SNR}(O)\sqrt{N}} \approx \frac{1}{\sqrt{pN}}
\end{equation}
To achieve a fixed relative precision, $\sigma_{\bar O}/\langle O \rangle \sim 1$, requires
\begin{equation}
    N \sim \frac{1}{p},
\end{equation}
so small signals $p \ll 1$ demand $N$ very large.

While quantum amplitude estimation~\cite{Brassard:2000xvp} can improve the asymptotic error scaling as $\mathcal{O}(1/N)$ instead of $\mathcal{O}(1/\sqrt{N})$, estimating exponentially small expectation values still requires an exponential circuit depth or shot overhead. 
Beyond pure states, let $\hat \rho$ be a density matrix with spectral decomposition:
\begin{equation}
    \hat{\rho} = \sum_i p_i | \psi_i \rangle \langle \psi_i |,
\end{equation}
where $p_i\ge 0 $, and $\sum_i p_i=1$. For any Hermitian $\hat O$,
\begin{equation}
\begin{aligned}
    \langle \hat{O} \rangle & = {\rm Tr}( \hat{\rho} \hat{O}) , \\
    {\rm Var}(O) & =  {\rm Tr}( \hat{\rho} \hat{O}^2) -  \left[{\rm Tr}( \hat{\rho} \hat{O})\right]^2.
\end{aligned}
\end{equation}
An ensemble of states $\{ |\psi_i \rangle \}$ drawn from probabilities $\{p_i\}$ produces the same variance and hence the same SNR scaling. 
The standard error with $N$ samples is again $\sqrt{{\rm Var}(\hat O)/N}$.

The quantities can be calculated in terms of the path integral formulation:
\begin{equation}
    \langle \hat O \rangle  = \frac{1}{Z} \int D\phi \, O(\phi) {\rm e}^{-S_E[\phi]}, \quad Z = \int D\phi \, {\rm e}^{-S_E[\phi]}.
\end{equation}
Here, $\phi$ denotes a field configuration and $S_E$ the Euclidean action. 
Classical Markov-chain Monte Carlo (e.g., Metropolis-Hastings~\cite{Metropolis:1953am, Hastings:1970aa} or Gibbs sampling~\cite{Geman:1984wrl}) generates configuration $\{\phi_i \}$ with probability density proportional to ${\rm e}^{-S_E[\phi]}$.
One then estimates
\begin{equation}
    \bar{O} = \frac{1}{N}  \sum_{i=1}^N O(\phi_i),
\end{equation}
with variance
\begin{equation}
     {\rm Var}_{\rm path}(O) =  \frac{1}{Z} \int D\phi \, O(\phi)^2 {\rm e}^{-S_E[\phi]} - \langle O \rangle^2,
\end{equation}
with standard error $\sigma_{\bar O} = \sqrt{{\rm Var}_{\rm path}(\hat {O})/N} $\footnote{Note that configurations should be decorrelated. While quantum computers are free from autocorrelation, one needs to prepare the state from scratch every time.}.
Up to discretization errors, this matches the variance of $\hat{O}$ using the density matrix formalism.

A canonical example of exponentially bad SNR in classical lattice QCD arises in nucleon two-point functions. Define
\begin{equation}
    C_N(t) = \langle 0 | {\mathcal N}(t) \bar{{\mathcal N}}(0) | 0 \rangle,
\end{equation}
where $\mathcal N$ is an interpolating operator with nucleon quantum numbers. The signal-to-noise ratio behaves as
\begin{equation}
    {\rm SNR}[C_N(t)] \sim \exp \left[-(m_N-\frac{3}{2}m_\pi)t \right],
\end{equation}
which decays exponentially in Euclidean time $t$ provided $m_N > 3m_\pi/2$ (as indeed it is)~\cite{Parisi:1983ae, Lepage:1989hd}. 
At large separations $t$, extracting $m_N$ from $C_N(t)$ thus requires an exponentially increasing number of Monte Carlo configurations $N_{\rm configs} \sim e^{2(m_N - \frac{3}{2}m_\pi)t}$.

To understand how the signal-to-noise problem manifests on a quantum computer in a physically realistic setting, consider Hamiltonian lattice gauge theory. 
A standard prototype is the 1+1D Schwinger model (QED$_2$) formulated with Kogut-Susskind staggered fermions on a spatial lattice of $N_s$ sites~\cite{Kogut:1974ag, Martinez:2016yna}. 
After integrating out the gauge degrees of freedom via Gauss's law and applying the Jordan-Wigner transformation, the Hamiltonian mapped to $N_s$ qubits takes the spin-chain form:
\begin{equation}
\begin{aligned}
    \hat{H} = & -w \sum_{n=1}^{N_s-1} \left( \sigma_n^+ \sigma_{n+1}^- + \sigma_n^- \sigma_{n+1}^+ \right) + \frac{m}{2} \sum_{n=1}^{N_s} (-1)^n \sigma_n^z \\
    & + \frac{g^2 a}{2} \sum_{n=1}^{N_s-1} \left[ \sum_{i=1}^n \frac{\sigma_i^z + (-1)^i}{2} \right]^2,
\end{aligned}
\end{equation}
where $w = 1/(2a)$ is the hopping parameter, $m$ is the bare fermion mass, $g$ is the gauge coupling, and $a$ is the lattice spacing.

Suppose we prepare the interacting vacuum $|\Omega\rangle$ and evaluate the real-time unequal-time meson correlator or pair-creation response:
\begin{equation}
    C(t) = \langle \Omega | \hat{M}(t) \hat{M}(0) | \Omega \rangle = \langle \Omega | {\rm e}^{i\hat{H}t} \hat{M} {\rm e}^{-i\hat{H}t} \hat{M} | \Omega \rangle,
\end{equation}
where $\hat{M}$ is a localized meson creation operator (e.g., $\hat{M} = \sigma_n^+ \sigma_{n+1}^- + \text{h.c.}$).
On a quantum processor, real-time unitary evolution ${\rm e}^{-i\hat{H}t}$ completely bypasses the sign problem that plagues classical Minkowski simulations.
However, as the localized excitation propagates and disperses into multi-particle continuum states, the real-time signal decays or oscillates with an envelope $|C(t)| \sim {\rm e}^{-\gamma t}$. 

When estimating $C(t)$ via projective measurements (e.g., decomposing $\hat{M}$ into Pauli strings $\hat{P}$ with eigenvalues $\pm 1$), Born's rule dictates that each single-shot measurement yields $\pm 1$, so the measurement variance satisfies:
\begin{equation}
    {\rm Var}(\hat{P}) = \langle \hat{P}^2 \rangle - \langle \hat{P}(t) \rangle^2 = 1 - \langle \hat{P}(t) \rangle^2 \xrightarrow{|C(t)| \to 0} 1.
\end{equation}
Consequently, the signal-to-noise ratio on the quantum device decays exponentially in real time:
\begin{equation}
    {\rm SNR}[C(t)] = \frac{|C(t)|}{\sqrt{{\rm Var}(\hat{P})}} \sim {\rm e}^{-\gamma t}.
\end{equation}
To achieve a target relative uncertainty $\sigma_{\bar C}/|C(t)| \le \epsilon$, the required number of quantum measurement shots scales as
\begin{equation}
    N_{\rm shots} \approx \frac{1}{\epsilon^2 \, {\rm SNR}^2} \sim \frac{1}{\epsilon^2} {\rm e}^{2\gamma t}.
\end{equation}
This demonstrates that even on an ideal fault-tolerant quantum computer where the sign problem is non-existent, Born-rule measurement stochasticity induces an exponential shot complexity identical in form to the classical Parisi-Lepage barrier.

Similarly, many thermodynamic observables and dynamical response functions are plagued by signal-to-noise problems.
Although quantum computers offer a promising avenue for simulating systems at finite density, where classical Monte Carlo methods are hindered by the sign problem, they do not inherently eliminate the signal-to-noise problem.
Estimating expectation values on a quantum device still involves stochastic sampling through repeated measurements, and observables with intrinsically large variances may demand a prohibitively large number of samples to achieve reliable precision.

\emph{Quantum control variates}---The control variates method is a well-established technique for variance reduction in classical Monte Carlo simulations~\cite{Kahn1953, Fieller1954, mcbook}.
The idea is to construct an auxiliary observable $f$ whose expectation value vanishes, $\langle f \rangle = 0$, and define a modified observable $\tilde O = O - f$ that retains the same expectation value as $O$ but may have a reduced variance.
Since $\langle \tilde O \rangle = \langle O \rangle$, this procedure yields an unbiased estimator with potentially improved statistical properties.
Recently, the effectiveness of this method has been significantly enhanced by combining it with machine learning models, which can be used to construct optimal or data-driven control variates in high-dimensional systems~\cite{Wan2018, Muller2020, grathwohl2018, Sun2021, Sun2023, Ott2023, Bedaque:2023ovz, Lawrence:2023cft, Oh:2025jaf, Oh:2025fpq, Siahkoohi:2026hna}.

A canonical recipe for constructing such an $f$ classically is Stein's method~\cite{stein1972bound}: for a distribution $p(\phi)\propto e^{-S(\phi)}$, integrating $\partial_\phi[g(\phi)e^{-S(\phi)}]$ over all $\phi$ gives $\langle g'(\phi) - g(\phi)S'(\phi)\rangle = 0$ for any function $g$, so $f(\phi) = g'(\phi) - g(\phi)S'(\phi)$ is a control variate for every choice of $g$.
This is a special case of a general principle: if a linear generator $L$ leaves the reference distribution stationary, then $\langle Lg \rangle = 0$ for every test function $g$ in its domain.
This classical approach can be generalized to the quantum setting by replacing the generator $L$, the test function $g$, and the reference distribution $p$ with an operator, an arbitrary Hermitian operator $A$, and the density matrix $\hat \rho$, respectively.
The central goal is again to identify an operator $\hat f$ such that $\langle \hat f \rangle = {\rm Tr}(\hat \rho \hat f)=0$, allowing one to replace a noisy observable $\hat O$ with a modified version $\hat O - \hat f$ with reduced variance.

To construct such operators in a principled way, consider the quantum canonical ensemble with density matrix $\hat \rho = {\rm e}^{-\beta \hat H}$. 
A useful observation is that for any operators $X$ and $A$, the following identity holds:
\begin{equation}
    {\rm Tr}([X, A \hat\rho]) = 0.
\end{equation}
If $X$ commutes with the Hamiltonian, $[X, \hat H]=0$, then it also commutes with the thermal density matrix, $[X, \hat \rho]=0$. In this case, the above identity simplifies to
\begin{equation}
    {\rm Tr}([X, A] \hat \rho) = 0,
\end{equation}
which implies that the operator
\begin{equation}
    \hat f=[X,A]  \label{Eq:QCV}
\end{equation}
has vanishing expectation value and may serve as a quantum control variate.
Here the commutator $[X,\cdot]$ plays the role of the generator $L$ above, with the stationarity condition $[X,\hat\rho]=0$ replacing invariance of the reference distribution.

To ensure that $\hat f$ is Hermitian, which is often required for measurement, one can restrict to the case where $X$ is anti-Hermitian and $A$ is Hermitian.
This choice preserves the Hermiticity of $\hat f$, since $([X, A])^\dagger = [A^\dagger, X^\dagger] = [A, -X] = [X, A]$.

This construction is universal for the off-diagonal part of the target operator: under mild conditions, it can represent any Hermitian operator with vanishing diagonal matrix elements in the eigenbasis of $X$.
Suppose $X$ is anti-Hermitian and non-degenerate, so that it admits a diagonalization $X = U^\dagger D U$, where $U$ is unitary and $D= {\rm diag}(i \lambda_1, \cdots, i\lambda_n) $ with $\lambda_i \neq \lambda_j$ for $i \neq j$.
Then, for any target control variate $\hat f$ whose diagonal vanishes in this basis, one can find a Hermitian operator $A$ such that $[X, A] = \hat f$.

To see this, define $\tilde A = U A U^\dagger$ and $\tilde f = U \hat f U^\dagger$, so that the equation becomes 
\begin{equation}
    [D, \tilde A] = \tilde f.
\end{equation}
In components, this reads 
\begin{equation}
    i(\lambda_i - \lambda_j) \tilde{A}_{ij} = \tilde{f}_{ij}.
    \label{Eq:CVComp}
\end{equation}
Since the eigenvalues $\lambda_i$ are all distinct by assumption, one can solve for the off-diagonal elements ${\tilde A}_{ij}$ uniquely for any given ${\tilde f}_{ij}$. 
Because $[D,\tilde A]$ vanishes identically on the diagonal in this basis, the construction spans exactly the operators with zero diagonal matrix elements in the eigenbasis of $X$ -- a proper subspace of the traceless operators, but one that already contains every purely off-diagonal (coherence) target relevant to canceling measurement noise between non-commuting observables.
For systems with non-degenerate Hamiltonians, a natural choice for $X$ is $-i \hat{H}$, ensuring commutation with $\hat \rho$ in thermal and stationary states.

The complementary diagonal sector admits an equally universal construction, but requires a different, discrete generator in place of the commutator, realizing the same stationarity principle directly on the populations rather than through $[X,\cdot]$.
The discrete quantum analogue acts on the spectrum of $\hat H$: let $\{|n\rangle\}$ be the energy eigenbasis, $n=1,\dots,N$, with populations $p_n = \langle n | \hat \rho | n \rangle \propto e^{-\beta E_n}$, and fix any cyclic ordering of the $N$ basis states.
Since the ratio $p_{n+1}/p_n = e^{-\beta(E_{n+1}-E_n)}$ depends only on the (measurable) energy gap and not on the partition function, the diagonal operator
\begin{equation}
    \hat{f}_{\rm diag} = \sum_n \left[ g_{n+1}\, e^{-\beta(E_{n+1}-E_n)} - g_n \right] |n\rangle\langle n|
    \label{Eq:SteinDiag}
\end{equation}
satisfies $\langle \hat{f}_{\rm diag} \rangle = 0$ for \emph{any} choice of the $N$ real parameters $\{g_n\}$, by telescoping around the cycle. The linear map from $\{g_n\}$ to the diagonal entries of $\hat f_{\rm diag}$ has a one-dimensional kernel (spanned by $g_n \propto 1/p_n$), so its $N$ parameters generate exactly the $(N-1)$-dimensional space of diagonal operators with vanishing thermal expectation value -- every such operator is reachable. Combined with the off-diagonal construction of Eq.~(\ref{Eq:CVComp}), the pair $\{[X,A], \hat{f}_{\rm diag}\}$ spans the full space of traceless Hermitian operators, giving a genuinely universal quantum control variate construction.

Given a parameterized operator $A(\bm{\theta}) = \sum_k \theta_k A_k$ where $\{A_k\}$ are Hermitian basis operators (e.g., Pauli strings or local gates), the modified observable $\tilde{O}(\bm{\theta}) = \hat{O} - [X, A(\bm{\theta})]$ has variance
\begin{equation}
    {\rm Var}(\tilde{O}(\bm{\theta})) = {\rm Var}(\hat{O}) - 2 \sum_k \theta_k v_k + \sum_{k,l} \theta_k \theta_l M_{kl},
\end{equation}
where
\begin{equation}
\begin{aligned}
    M_{kl} &= \frac{1}{2} \langle \{ [X, A_k], [X, A_l] \} \rangle, \\
    v_k &= \frac{1}{2} \langle \{ \hat{O}, [X, A_k] \} \rangle.
\end{aligned}
\end{equation}
Because the covariance matrix $M$ is positive semi-definite, finding the optimal parameters $\bm{\theta}^* = M^{-1} \bm{v}$ reduces to solving a standard convex linear system, establishing a direct quantum analogue of classical optimal control variates.

\emph{Illustrative Examples}---To provide concrete intuition into how quantum control variates operate, we examine three representative examples: a minimal single-qubit model, a continuous-variable system connecting to the quantum Virial theorem, and a multi-qubit spin model with symmetry constraints.

\textbf{1. Single-qubit toy model.} Consider a single qubit governed by the Hamiltonian $\hat{H} = \frac{\omega}{2} \sigma_z$ in a thermal state $\hat{\rho} = {\rm e}^{-\beta \hat{H}}/Z = \frac{1}{2}\left(I - \tanh(\beta\omega/2) \sigma_z \right)$.
Choose the generator $X = -i \hat{H} = -i \frac{\omega}{2} \sigma_z$, which trivially satisfies $[X, \hat{\rho}] = 0$. 

Let $A = \alpha \sigma_y$ with real parameter $\alpha$. The control variate is
\begin{equation}
    \hat{f} = [X, A] = \left[ -i \frac{\omega}{2} \sigma_z, \alpha \sigma_y \right] = -\alpha \omega \sigma_x.
\end{equation}
Since $\langle \sigma_x \rangle = {\rm Tr}(\hat{\rho} \sigma_x) = 0$, we have $\langle \hat{f} \rangle = 0$ for any $\alpha$.
Now suppose we are interested in estimating an observable with transverse fluctuations, $\hat{O} = \sigma_z + \delta \sigma_x$, where the true expectation value is $\langle \hat{O} \rangle = \langle \sigma_z \rangle = -\tanh(\beta\omega/2)$.
The unmitigated variance is ${\rm Var}(\hat{O}) = {\rm Var}(\sigma_z) + \delta^2$. The modified observable is
\begin{equation}
    \tilde{O} = \hat{O} - \hat{f} = \sigma_z + (\delta + \alpha \omega) \sigma_x.
\end{equation}
Setting $\alpha^* = -\delta / \omega$ completely removes the transverse fluctuation $\delta \sigma_x$, yielding $\tilde{O} = \sigma_z$ and reducing the variance to ${\rm Var}(\sigma_z)$, eliminating the extrinsic statistical noise $\delta^2$.

\textbf{2. Quantum Virial theorem.} A compelling physical realization arises in continuous-variable quantum systems. Consider a particle with Hamiltonian $\hat{H} = \hat{T} + \hat{V} = \frac{\hat{p}^2}{2m} + V(\hat{x})$. Choosing $X = -i \hat{H}$ and $A = \frac{1}{2}(\hat{x}\hat{p} + \hat{p}\hat{x})$, which generates spatial dilatations, the control variate operator is
\begin{equation}
    \hat{f} = -i [\hat{H}, A] = \hat{x} V'(\hat{x}) - 2\hat{T}.
\end{equation}
The condition $\langle \hat{f} \rangle = 0$ in any stationary or thermal state $\hat{\rho}$ is precisely the \emph{quantum Virial theorem}: $2\langle \hat{T} \rangle = \langle \hat{x} V'(\hat{x}) \rangle$. 

For a harmonic oscillator $V(\hat{x}) = \frac{1}{2}m\omega^2 \hat{x}^2$, we obtain $\hat{f} = 2(\hat{V} - \hat{T})$. 
When estimating the kinetic energy $\hat{O} = \hat{T}$, single-shot projective measurements exhibit non-zero variance ${\rm Var}(\hat{T}) > 0$ even in energy eigenstates.
Defining the modified estimator $\tilde{T} = \hat{T} - c \hat{f} = (1+2c)\hat{T} - 2c\hat{V}$, the choice $c^* = -1/4$ gives $\tilde{T} = \frac{1}{2}(\hat{T} + \hat{V}) = \frac{1}{2}\hat{H}$.
In any energy eigenstate $|n\rangle$, ${\rm Var}(\tilde{T}) = \frac{1}{4}{\rm Var}(\hat{H}) = 0$, achieving exact zero-variance estimation.

\textbf{3. Symmetry-induced control variates.} In many-body systems with conserved global charges, symmetries provide a natural reservoir of control variates: any conserved generator automatically satisfies the stationarity condition required above.
Suppose the Hamiltonian commutes with a symmetry generator $\hat{Q}$, $[Q, \hat{H}] = 0$, so that $[Q, \hat{\rho}] = 0$ in thermal equilibrium---precisely this stationarity condition. Choosing $X = -i \hat{Q}$, any operator $\hat{f}_Q = [-i \hat{Q}, A]$ is a zero-mean control variate for arbitrary Hermitian $A$.

As an example, consider a two-qubit exchange model $\hat{H} = J (\sigma_1^x \sigma_2^x + \sigma_1^y \sigma_2^y) + h (\sigma_1^z + \sigma_2^z)$ possessing $U(1)$ spin-rotational symmetry generated by $\hat{Q} = \hat{J}_z = \frac{1}{2}(\sigma_1^z + \sigma_2^z)$. Taking $X = -i \hat{J}_z$ and $A = \sigma_1^x \sigma_2^y$, we find
\begin{equation}
    \hat{f}_Q = [-i \hat{J}_z, \sigma_1^x \sigma_2^y] = \sigma_1^y \sigma_2^y - \sigma_1^x \sigma_2^x.
\end{equation}
Because $\langle \hat{f}_Q \rangle = 0$ identically due to the continuous rotation symmetry $\langle \sigma_1^x \sigma_2^x \rangle = \langle \sigma_1^y \sigma_2^y \rangle$, $\hat{f}_Q$ can be used to balance measurement noise between different Pauli components of the interaction energy without biasing the expectation value.
Note that $\hat{J}_z$ is degenerate here, so this example does not by itself exercise the general non-degenerate construction of Eq.~(\ref{Eq:CVComp}).
A complementary case with a manifestly non-degenerate generator is a single bosonic mode with $U(1)$ particle-number symmetry, e.g.\ a Kerr oscillator $\hat{H} = \omega \hat{n} + \chi\, \hat{n}(\hat{n}-1)$ with $\hat{n}=a^\dagger a$, whose eigenvalues $0,1,2,\dots$ are all distinct.
Choosing $X=-i\hat{n}$ and $A=\hat{p}=\frac{i}{\sqrt2}(a^\dagger - a)$, the identity $[\hat{n},\hat{p}]=i\hat{x}$ gives
\begin{equation}
    \hat{f}_Q = [-i\hat{n}, \hat{p}] = \hat{x}, \qquad \hat{x}=\tfrac{1}{\sqrt2}(a+a^\dagger).
\end{equation}
Any number-conserving state ($[\hat{n},\hat\rho]=0$, automatic whenever $\hat H$ is diagonal in $\hat n$) has no coherence between Fock states, so $\langle \hat{x} \rangle = 0$ identically -- a phase-symmetric state carries no mean field amplitude.
Because $\hat n$ is non-degenerate, this example saturates the construction of Eq.~(\ref{Eq:CVComp}) exactly.
As a concrete use, suppose a number-resolving detector suffers a small residual homodyne cross-talk, $\hat{O}=\hat n + \delta \hat x$; the choice $c^\star=\delta$ in $\tilde{O}=\hat{O}-c\hat{f}_Q$ exactly cancels the spurious quadrature contamination, giving $\tilde{O}=\hat n$ and removing the extrinsic variance $\delta^2 {\rm Var}(\hat x)$.
More generally, parameterizing $A$ via a variational quantum circuit and enforcing known physical symmetries allows systematic variance reduction across complex many-body quantum simulations.

\begin{figure}[t]
    \centering
    \includegraphics[width=\linewidth]{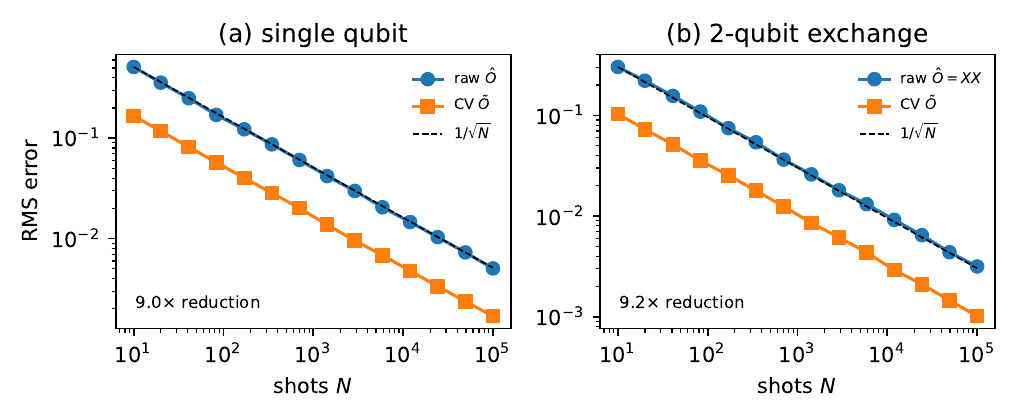}
    \caption{RMS estimation error vs.\ number of shots $N$, with and without the quantum control variate, for (a) the single-qubit toy model and (b) the two-qubit $U(1)$-symmetric exchange model.
    Both estimators follow the expected $1/\sqrt{N}$ scaling (dashed line); the control-variate estimator $\tilde O$ has uniformly lower error at fixed $N$.}
    \label{fig:cv_demo}
\end{figure}

\textbf{4. Numerical validation.} We confirm these results quantitatively by simulating both the single-qubit model and the $U(1)$-symmetric two-qubit exchange model exactly, using finite-shot Born-rule sampling to emulate real projective measurements (Fig.~\ref{fig:cv_demo}).
For the single-qubit model ($\omega=1.0$, $\beta=2.5$, $\delta=1.5$), the optimal control variate reduces the estimator variance by a factor of $9.0$.
For the two-qubit exchange model ($J=1$, $h=2.0$, $\beta=1.0$), $\hat\sigma_1^x\hat\sigma_2^x$ and $\hat\sigma_1^y\hat\sigma_2^y$ commute and are therefore read out jointly from a single measurement circuit, so the symmetry-induced control variate $\hat f_Q$ reduces the variance of $\langle \hat\sigma_1^x\hat\sigma_2^x\rangle$ by a factor of $9.2$ at no additional measurement cost.
These factors are finite rather than exact because each example uses a single control-variate direction on a thermal (mixed) state: the residual variance is the component of ${\rm Var}(O)$ uncorrelated with $\hat f$, in contrast to the pure-eigenstate zero-variance limit of Example 2.
Both estimators retain the expected $1/\sqrt{N}$ shot-noise scaling, confirming that the correction is a genuine, unbiased variance reduction rather than a change in convergence rate.

\emph{Discussion}---The central message of this paper is that avoiding the sign problem is not the same as avoiding statistical noise.
Fault-tolerant quantum computers implement unitary real-time evolution exactly, but the Born rule still forces expectation values to be reconstructed from finite, stochastic measurement outcomes, and the associated signal-to-noise ratio can decay exponentially in the same physical regimes---long Euclidean or real time, low temperature, large volume---that plague classical Monte Carlo.
Because sign-problem avoidance is often mistaken for a proxy for algorithmic efficiency, this point is easy to overlook; our analysis and numerical examples show that variance reduction must be treated as a separate, equally essential ingredient of any practical quantum simulation strategy.

We proposed a quantum generalization of Stein control variates construction, built from operator commutators $[X,A]$ together with a discrete diagonal counterpart, and verified both its underlying identities and its practical effect numerically: the exactly solvable single-qubit and $U(1)$-symmetric two-qubit models exhibit unbiased variance reductions of order unity using only classical post-processing of existing measurement data.
For larger, non-integrable systems, however, computing the optimal $\bm\theta^*=M^{-1}\bm v$ directly requires evaluating $O(k^2)$ commutator expectation values, which itself becomes a resource bottleneck.
A promising direction is to instead parameterize $A$ via a variational quantum circuit and adjust its parameters through a hybrid classical-quantum optimization loop---analogous to VQE---that directly minimizes the measured variance rather than solving the linear system explicitly; we leave a detailed study of such variationally optimized control variates to future work.

More broadly, realizing the promised advantages of quantum simulation for many-body physics will require systematically combining structural noise-mitigation techniques, such as the control variates discussed here, with complementary strategies like quantum amplitude estimation.
Treating variance reduction as a first-class algorithmic problem, rather than an afterthought, is necessary for quantum computation to deliver on its promise as a tool for nonperturbative many-body physics.

\emph{Acknowledgments}---The author thanks Tom Cohen for helpful discussions in the early stages of this work.
This work was supported in part by the U.S. Department of Energy, Office of Nuclear Physics under Award Number DE-FG02-93ER40762.

\bibliography{refs.bib}

\end{document}